\documentclass[aps,prb,epsfig]{revtex4}
\usepackage{epsfig}
\usepackage{amsmath}
\usepackage{amssymb}
\begin{document}
\title{Dynamical and point symmetry of the Kondo effect 
in triangular quantum dot}
\author{$^1$T. Kuzmenko, $^1$K. Kikoin, and $^{1,2}$Y. Avishai}
\affiliation{$^1$Department of Physics, Ben-Gurion University of
the
Negev, Beer-Shea 84105, Israel\\
$^2$ Ilse Katz Center for Nano-Technology,  Ben-Gurion University
of the Negev, Beer-Sheva 84105, Israel\\
}
\date{\today}
\begin{abstract}
In this work we concentrate on the {\it point symmetry} of triangular triple
quantum dot
and its interplay with the {\it spin rotation symmetry} in the
context of Kondo tunneling through this kind of artificial molecule.
 A fully symmetric triangular triple quantum dot  is considered, 
consisting of three identical puddles with the same individual 
properties (energy levels and Coulomb blockade parameters) and
inter-dot coupling (tunnel amplitudes and electrostatic
interaction).  The underlying Kondo physics is determined by
the product of a discrete rotation symmetry group in real space 
and a continuous rotation symmetry in spin 
space. These symmetries are 
reflected in the resulting exchange hamiltonian which 
naturally involves spin and orbital degrees of freedom. 
The ensuing poor-man scaling equations are solved and the Kondo temperature 
is calculated.
\end{abstract}
\maketitle

\section{Introduction}

The analogy between complex quantum dots and real molecules was
recognized both by experimentalists and theoreticians at early
stage of studies of these artificial nanoobjects (see, e.g.,
\cite{dqd,analog}). Soon after the discovery of the Kondo effect
in tunneling through quantum dots (QD) \cite{KKK}, it was
recognized that this phenomenon may be realized in tunneling
through complex quantum dots consisting of two or three electron
puddles (double ant triple quantum dots) \cite{DD,KAv,Hofsh,KuKA}.
Tunneling through these "artificial linear  molecules" has been
studied mainly in serial and parallel configurations, where the
dots are ordered linearly either parallel or perpendicular to the
metallic leads. Meanwhile, modern experimental methods allow also
fabrication of quantum dots in a triangular geometry. Triangular
triple quantum dot (TTQD) was considered theoretically
\cite{stop1} and realized experimentally very recently,
\cite{stop2} in order to demonstrate the ratchet effect in single
electron tunneling. To achieve this effect the authors proposed a
configuration, where two of the three puddles are coupled in
series with the leads (source and drain), while the third one has
a tunnel contact with one of its counterparts and only a
capacitive coupling with the other.

In this paper we concentrate on the {\it point symmetry} of TTQD
and its interplay with the {\it spin rotation symmetry} in a
context of Kondo tunneling through this artificial molecule.
Indeed, the generic feature of Kondo effect is the involvement of
internal degrees of freedom of localized "scatterer" in the
interaction with continuum of electron-hole pair excitations in
the Fermi sea of conduction electrons. These are spin degrees of
freedom in conventional Kondo effect, although in some cases the
role of pseudospin may be played by configuration quantum numbers,
like in two-level systems and related objects. \cite{Coza} TTQD
may be considered as a specific Kondo object, where both spin and
configuration excitations are involved in cotunneling on equal
footing.

\begin{figure}[htb]
\centering
\includegraphics[width=130mm,height=100mm,angle=0]{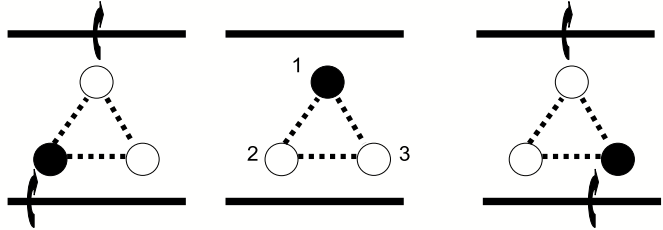}
\caption{Triangular triple quantum dot (TTQD) in a contact with
source (s) and drain (d) leads. Clockwise (c) and anti-clockwise
(a) "rotation" of TTQD due to cotunneling through the channels
$V_{s3}$ and $V_{s2}$, respectively.}\label{figTQD}
\end{figure}

To demonstrate this interplay, we consider a fully symmetric TTQD
consisting of three identical puddles with the same individual
properties (energy levels and Coulomb blockade parameters) and
inter-dot coupling (tunnel amplitudes and electrostatic
interaction. Like in the above mentioned triangular ratchet
\cite{stop1,stop2}, we assume that the TTQD in the ground state is
occupied by one electron and Coulomb blockade is strong enough to
completely suppress double occupancy of any valley $j=1,2,3$. This
means that the only mechanism of electron transfer through TTQD is
cotunneling, where one electron leaves the valley $j$ for metallic
leads, whereas another electron tunnels from reservoir to the same
valley $j$ or to another valley $l$. In the former case only the
spin reversal is possible, whereas in the latter case not only the
spin is affected but also the TTQD is effectively "rotated" either
clockwise or anti-clockwise (see Fig. \ref{figTQD}).

Discrete rotation in real space and continuous rotation in spin
space may be described in terms of group theory. The group
$C_{3v}$ characterizes the symmetry of triangle, and the group
$SU(2)$ describes the symmetry of spin 1/2. So the total symmetry
of TTQD is determined by direct product of these two groups. One
may use an equivalent language of permutation group $P_3$ for
description of configuration of TTQD with an electron occupying
one of three possible positions in its wells.

\section{Hamiltonian of artificial triangular molecule}

A symmetric TTQD in a contact with source and drain leads (Fig.
\ref{figTQD}) is described by the Hamiltonian
\begin{equation}
H=H_{d}+H_{lead}+H_t. \label{H-sys}
\end{equation}
The first term $H_d$ is the Hamiltonian of the isolated TQD,
\begin{eqnarray}
H_{d}=\epsilon\sum_{j=1}^3\sum_{\sigma}d^\dagger_{j
\sigma}d_{j\sigma }+Q\sum_{j}n_{j\uparrow}n_{j\downarrow}
+Q'\sum_{\langle jl\rangle}\sum_\sigma n_{j\sigma}n_{l\sigma'}
+W\sum_{\langle jl\rangle}\sum_{\sigma}(d^\dagger_{j
\sigma}d_{l\sigma }+H.c.), \label{H-dot}
\end{eqnarray}
 where ${\sigma}=\uparrow,\downarrow$ is the spin index, $\langle jl \rangle=
\langle12\rangle,\langle23\rangle,\langle31\rangle$. $Q$ and $Q'$
are intra-dot and inter-dot
 Coulomb blockade parameters, $W_{jl}$ are inter-dot tunneling
 parameters.
 The second term $H_{lead}$
describes the electrons in the source $(s)$ and drain $(d)$
electrodes,
\begin{eqnarray}
H_{lead}&=&\sum_{k\sigma}\sum_{b=s,d}\epsilon_{kb}c^\dagger_{kb
\sigma}c_{kb\sigma}. \label{H-l}
\end{eqnarray}
The last term $H_t$ is the tunneling Hamiltonian
\begin{eqnarray}
H_t=\sum_{k\sigma}\left(\sum_{j=2,3} V_{sj}c^\dagger_{ksj\sigma}
d_{j\sigma}+V_sd^\dagger_{kd\sigma }d_{1\sigma}
+H.c.\right).\label{H-tun}
\end{eqnarray}
which describes a configuration with two-channel tunneling between
the TTQD and the source and single-channel tunneling between the
TTQD and the drain. If all three channels are equivalent,
\begin{equation}\label{onechan}
V_{d}=V_{s2}= V_{s3}\equiv V,
\end{equation}
the tunnel contacts preserve the rotation/permutation symmetry
of TTQD. If all tunnel constants are different, this symmetry is
completely destroyed by tunneling. In the intermediate case
\begin{equation}\label{oneax}
V_{d}\neq V_{s2}= V_{s3}\equiv V_s,
\end{equation}
the system preserves one mirror reflection axis $(2
\leftrightarrow 3)$. The Kondo tunneling regime will be analyzed
both in cases (\ref{onechan}) and (\ref{oneax}).

The Hamiltonian $H$ (\ref{H-sys}) is easily diagonalized by using
the following set of basis functions $|j\sigma\rangle=
d^\dag_{j\sigma}|0\rangle$ ($j=1,2,3$). The eigenfunctions of
symmetric TTQD in this basis are
\begin{eqnarray}
\Psi_{\sigma,A_1}&=&
\frac{1}{\sqrt{3}}(|1\sigma\rangle+|2{\sigma}\rangle+|3{\sigma}\rangle)\label{A1},\\
\Psi_{\sigma,E(+)}&=&
\frac{1}{\sqrt{3}}(|1\sigma\rangle+e^{2i\varphi}|2{\sigma}\rangle
    +e^{i\varphi}|3{\sigma}\rangle)\nonumber,\\
\Psi_{\sigma,E(-)}&=&
\frac{1}{\sqrt{3}}(|1\sigma\rangle+e^{i\varphi}|2{\sigma}\rangle
+e^{2i\varphi}|3{\sigma}\rangle)\nonumber
\end{eqnarray}
with $\varphi=2\pi/3$. Here $A_1$ and $E$ are two irreducible
representations of the group $C_{3v}$
 The spin states  with ${\cal N}=1$
are spin doublets ($D$), so the Hamiltonian of isolated TTQD in
this charge sector has six eigenstates $|DA\rangle, |DE\rangle$ .
The single electron energies are
\begin{equation}\label{energy-1}
E_{DA_1}=\epsilon+2W, ~~~~~~~~~~~E_{DE}=\epsilon-W.
\end{equation}
First of these energy levels is a conventional spin doublet with
fully symmetrical "orbital" wave function $\Psi_{\sigma,A_1}$. The
second one is doubly degenerate both in spin and orbital quantum
number.

A peculiar feature of the three-site configuration is an {\it
explicit dependence of the order of levels} in spin multiplet on
the sign of tunnel integral $W$. The ground state is a spin
doublet $E_{DA_1}$, provided $W<0$. In case of  $W>0$ the lowest
level is the orbital doublet $E_{DE}$.  The orbital degeneracy of
the states $E(\pm)$ is a manifestation of rotation/permutation
degrees of freedom of TTQD illustrated by Fig. 1. These discrete
rotations are explicitly involved in Kondo tunneling.

\section{Kondo tunneling through partially occupied TTQD}

A contact between symmetric TTQD and leads in the two-terminal
geometry preserves  rotation symmetry $C_{3v} (P_3)$ unless the
tunnel Hamiltonian $H_t$ has lower symmetry than TTQD. In any case
it is useful to re-expand the tunnel Hamiltonian in partial waves,
which respect the "point symmetry" of the Hamiltonian $H$. Such
ansatz  is known in the theory of Kondo effect in 3D metal, where
the spherical partial  wave representation for Bloch electrons in
conduction band was introduced in Ref. [\onlinecite{CqS}]. Later
on, this approach was extended by many authors to the case of
point crystal group representation (see, e.g., the review
[\onlinecite{Coza}]). Here we also meet the situation where the
triangular quantum dot imposes its point symmetry on the continuum
of electron states in the leads under certain conditions.

If the condition (\ref{onechan}) is satisfied, the point symmetry
of the device as a whole is still $C_{3v}$, and the band states
may be re-expanded as
\begin{eqnarray}
c_{A,{\bf{k}},\sigma}
 &=&
 \frac{1}{\sqrt{3}}
 \left(
      c_{d,{\bf{k}},\sigma}+
      c_{s_2,{\bf{k}},\sigma}+
      c_{s_3,{\bf{k}},\sigma}
 \right). \label{partw}\\
 c_{E(+),{\bf{k}},\sigma}
 &=&
 \frac{1}{\sqrt{3}}
 \left(
      c_{d,{\bf{k}},\sigma}+
      e^{2i\varphi}c_{s_2,{\bf{k}},\sigma}+
      e^{i\varphi}c_{s_3,{\bf{k}},\sigma}
 \right),
 \nonumber \\
 c_{E(-),{\bf{k}},\sigma}
 &=&
 \frac{1}{\sqrt{3}}
 \left(
      c_{d,{\bf{k}},\sigma}+
      e^{i\varphi}c_{s_2,{\bf{k}},\sigma}+
      e^{2i\varphi}c_{s_3,{\bf{k}},\sigma}
 \right),
 \nonumber
\end{eqnarray}
which has the same angular dependence as (\ref{A1}).

The Anderson Hamiltonian $H$ rewritten in these variables may be
expressed by means of Hubbard operators
$X^{\lambda\lambda'}=|\lambda\rangle\langle \lambda'|$, with
$\lambda=0,\gamma,\Gamma,\Lambda$:
\begin{eqnarray}
H&=& E_0 X^{00} + \sum_\lambda E_\lambda X^{\lambda\lambda}+
\sum_\Gamma E_\Gamma X^{\Gamma\Gamma}+
\sum_{k\sigma}\sum_{k\gamma}
\varepsilon_{k}n_{ k \gamma} \label{hah}\\
&+&\sum_{\gamma} \left[V^{\gamma 0} X^{\gamma 0}
c_{\gamma}+\sum_{\Gamma\gamma\gamma'}
V^{\gamma\Gamma}c^\dagger_{\gamma}X^{\gamma\Gamma}+h.c.\right]\nonumber
\end{eqnarray}
Here $|0\rangle$ stands for an empty TTQD,
$|\gamma\rangle=|DA_1\rangle,~|DE\rangle$ belong to the single
electron charge sector, and $|\Gamma\rangle$ are the eigenvectors
of two-electron states. The eigenstates $E_\Gamma$ for ${\cal
N}=2$ are
\begin{eqnarray}
E_{SA_1} &=&{\epsilon_2}+2W- \frac{8W^2}{Q},
\nonumber\\
E_{TE} &=& {\epsilon_2} + W, \nonumber \\
E_{SE} &=& {\epsilon_2}- W - \frac{2W^2}{Q},
\nonumber\\
E_{TA_2} &=& {\epsilon_2} -2W  .\label{energy-n}
\end{eqnarray}
Here $\epsilon_2=2\epsilon+Q'$, indices $S,T$ denote spin singlet
and spin triplet configurations of two electrons in TTQD, and the
inequality $W\ll Q$ is used explicitly. The irreducible
representation $A_2$ contains two-electron eigenfunction, which is
odd with respect to permutations $j \leftrightarrow l$. Like in
the singly occupied TTQD, the level ordering is sensitive to the
sign of tunnel integral $W$.

The tunnel matrix elements are redefined accordingly.
$$
    V^{0\gamma}=\langle
\psi_{c,\gamma}|H_t|\Psi_\gamma\rangle, ~~~
    V^{\gamma\Gamma}=\langle
\psi_{c,\gamma'},\Psi_{\gamma}|H_t|\Psi_\Gamma\rangle.
$$
The matrix elements intermixing the charge sectors ${\cal N}=1,2$,
add conduction electron $\gamma'$ to the dot electron $\gamma$,
and the resulting two-electron states $\Gamma$ arise in accordance
with Clebsch-Gordan coefficients both in spin and orbital
subspaces $ {\cal C}_{\rho\rho'}^{\rho{''}} {\cal
C}_{\sigma\sigma'}^M $ where $\rho=A, E(\pm)$, $M=S,T$.

We describe Kondo tunneling by means of Haldane-Anderson
renormalization group (RG) approach \cite{Andrg,Hald}. According
to this procedure the parameters of the original Hamiltonian are
renormalized in the course of rescaling of the energy ${\cal D}$
characterizing the width of conduction electron continuum in the
leads from original value ${\cal D}_0$ to the energy $T_K$, which
characterizes Kondo correlations. The latter is found from the
solution of scaling equations. First, the energy levels $E_\gamma$
are renormalized by the virtual excitations of the states
$|0\rangle$ and $|\Gamma\rangle$. The flow equations have the form
\begin{eqnarray}
 E_{A}({\cal D})&=&
 \epsilon+2W- \Delta \ln\left({\cal D}_0/{\cal D}\right),
 \nonumber\\
 E_{E}({\cal D})&= &\epsilon-W- \Delta
 \ln\left({\cal D}_0/{\cal D}\right),\label{ren2}
 \end{eqnarray}
where $\Delta \sim \rho_0 V^2$ is the tunneling rate, $\rho_0$ is
the density of electron states in the leads. This scaling ends at
${\cal D} \to \bar {\cal D}$, with $\bar {\cal D}$ defined as
$$E_\gamma(\bar {\cal D})\sim \bar {\cal D}$$
(Schrieffer-Wolff limit) \cite{Hald}. Besides, the
Haldane-Anderson procedure generates effective two-particle
vertices $J_\gamma$, which describe effective exchange interaction
between dot and lead electrons. Various cotunneling processes
contribute to this interaction. In conventional Kondo problem the
effective low-energy exchange Hamiltonian has the form $J {\bf S}
\cdot {\bf s}$, where ${\bf S}$ and ${\bf s}$ are the spin
operators for dot and lead electrons, respectively \cite{Andrg}.
Here we are in a position, where the low-energy states of the
quantum dot are represented by a multiplet containing both spin
and orbital indices. In this case the form of effective
Hamiltonian is predetermined by a {\it dynamical symmetry} of the
Hamiltonian $H_d$ \cite{KAv,KKAv}.

The dynamical symmetry group of a given Hamiltonian is determined
not only by the operators, which leave this Hamiltonian invariant
but also by the operators describing transitions between different
energy levels of the multiplet. It follows from this definition
that the dynamical symmetry may vary depending on the set of
energy levels which fall into an actual energy interval. In
particular, it changes in the process of rescaling the energy
interval in the Anderson's RG procedure \cite{Andrg}. If the
condition $\bar {\cal D} \sim 3W$ is satisfied, then the dynamical
symmetry is determined by the whole multiplet $E_\gamma$. The
dynamical symmetry group, which describes all possible transitions
within the set $\{DA_1,DE(\pm)\}$ is $SU(6)$. Among 35 generators
of this group are nine spin vectors $\bf S_{\rho\rho'}$ with
projections defined as
\begin{eqnarray}
S^{+}_{\rho\rho'}=X^{\uparrow\rho,\downarrow\rho'},~~~
S^{-}_{\rho\rho'}=X^{\downarrow\rho,\uparrow\rho'},\label{spinrho}\\
S^{z}_{\rho\rho'}=\frac{1}{2}(X^{\uparrow\rho,\uparrow\rho'}-
X^{\downarrow\rho,\downarrow\rho'}).\nonumber
\end{eqnarray}
The rest 8 operators describe the permutation degrees of freedom
of TTQD. In the process of further reduction of the energy scale
${\cal D}$ the highest of two levels $E_\gamma$ is quenched, and
eventual symmetry of Kondo effect depends on the sign of tunneling
amplitude $W$.

In case of $W<0$ the permutation degrees
 of freedom are quenched at low-energy scale.
The only vector, which is involved in the Kondo cotunneling
through TTQD is the spin ${\bf S}_{AA}$ defined in Eq.
(\ref{spinrho}). The lead states are still classified in
accordance with the point symmetry of the system. As a result the
SW Hamiltonian has the form
\begin{eqnarray}\label{SWA}
 H_{SW} =
 J_E\left({\bf S}\cdot{\bf s}_{+}+
 {\bf S}\cdot{\bf s}_{-}\right)+
 J_{A}{\bf S}\cdot{\bf s}_A,
\end{eqnarray}
(the subindex $A$ of spin operator of TQD is omitted). The
exchange vertices $J_\rho$ are
\begin{eqnarray}\label{coupl-Ja}
 J_E &=&-\frac{2V^2}{3}
   \left(\frac{1}{\epsilon + Q'-\epsilon_F}
   -\frac{1}{\epsilon+Q- \epsilon_F}\right), \label{new-J}\\
  J_{A}&=& \frac{2V^2}{3}
   \left(\frac{3}{\epsilon_F
   -\epsilon}+
   \frac{1}{\epsilon+Q-\epsilon_F}
   +\frac{2}{\epsilon+Q'-\epsilon_F}\right).\nonumber
\end{eqnarray}
The constant $J_A$ has antiferromagnetic sign like in conventional
SW case, whereas the constant $J_E$ is negative due to the
inequality $Q\gg Q'$.

Formally, the Hamiltonian (\ref{SWA}) describes three-channel
cotunneling.\cite{NB} However, two of three available exchange
channels in the Hamiltonian (\ref{SWA}) are irrelevant for Kondo
cotunneling, because the coupling constant $J_E$ is negative
(ferromagnetic). As a result the conventional Kondo regime arises
in $DA$ channel with Kondo temperature
\begin{eqnarray}
T_{K}^{(A)}={\bar D}\exp{\left\{-\frac{1}{\rho_0
J_A}\right\}}.\label{TKA}
\end{eqnarray}
where $\rho_0$ is the density of electron states in the leads,
which is assumed to be the same for all channels. In case of $W>0$
the doublet $E_{DA}$ is quenched at $\bar D < W$, and the Kondo
effect is defined by the tunneling through TTQD in the state
$|D,E\rangle$ whose symmetry is $SU(4)$. Similar situation was
observed for a double quantum dot in series geometry.
\cite{Bord03} In that case the pseudospin variable describes two
possible occupations of an electron in a double dot. The 15
generators of this group bunch in four spin vectors $\bf
S_{\rho\rho'}$ with $\rho=E(\pm))$ and one pseudospin vector
${\boldsymbol {\cal T}}$ defined as
\begin{eqnarray}
&&{\cal T}^{+}=\sum_\sigma X^{\sigma+,\sigma
-},~~{\cal T}^{-}=\sum_\sigma X^{\sigma -,\sigma +}\nonumber\\
&& {\cal T}^z=\frac{1}{2}\sum_\sigma\left( X^{\sigma +,\sigma +}-
X^{\sigma -,\sigma -}\right).
\end{eqnarray}
Here the indices $\pm$ stand for "orbital" indices $E({\pm})$ for
the sake of brevity.

 Due to high degeneracy of the ground state of TTQD, the
effective SW Hamiltonian acquires quite complicated form,
\begin{eqnarray}
H_{SW}&=&J_1({\bf S}_{+}\cdot {\bf s}_{+}+{\bf S}_{-}\cdot {\bf
s}_{-})+J_2({\bf S}_{+}\cdot {\bf s}_{-}+{\bf S}_{-}\cdot {\bf
s}_{+})\nonumber\\
&+&J_3({\bf S}_{+}+{\bf S}_{-})\cdot {\bf s}_{A}+J_4({\bf
S}_{+-}\cdot {\bf s}_{-+}+{\bf S}_{-+}\cdot {\bf
s}_{+-})\nonumber\\
&+&J_5({\bf S}_{+-}\cdot ({\bf s}_{A-}+{\bf s}_{+A})+{\bf
S}_{-+}\cdot ({\bf s}_{A+}+{\bf s}_{-A}))\nonumber \\
&+& J_6 \boldsymbol {\cal T}\cdot{\boldsymbol \tau} ,\label{SWE}
\end{eqnarray}
where
\begin{eqnarray*}
  \tau_z &=&
  \frac{1}{2}\sum_\sigma
  \left(
       c^{\dag}_{\sigma_{+}}
       c_{\sigma_{+}}
       -
       c^{\dag}_{\sigma_{-}}
       c_{\sigma_{-}}
  \right),
  \\
  \tau^{+} &=&\sum_\sigma
         c^{\dag}_{\sigma_{+}}c_{\sigma_{-}}
         ,
  \ \ \ \ \
  \tau^{-}=\left(\tau^{+}\right)^{\dag},
\end{eqnarray*}
and the coupling constants $J_{1-6}$ are
\begin{eqnarray}
 J_1&=&J_4=\frac{2V^2}{3}
   \left(\frac{3}{\epsilon_F
   -\epsilon} \right.\nonumber\\
   &+&\left.\frac{1}{\epsilon+Q -\epsilon_F}
   +\frac{2}{\epsilon+Q' -\epsilon_F}\right).\nonumber\\
J_2&=&J_3=J_5
\nonumber\\
 &=&-\frac{2V^2}{3}
   \left(\frac{1}{\epsilon+Q'-\epsilon_F}
   -\frac{1}{\epsilon+Q-\epsilon_F}\right).\nonumber\\
J_6&=&
 \frac{V^2}{{\epsilon_F
   -\epsilon}}+
 \frac{V^2}{\epsilon+Q'-\epsilon_F}.\label{J-E}
\end{eqnarray}

Both spin and "orbital" degrees of freedom of TTQD are involved in
effective Kondo tunneling. The indirect exchange coupling
constants arise as combinations of cotunneling processes with
virtual excitation of states with zero and two electrons. It is
worth mentioning that the rotation $C_{3v}$ symmetry in a system
dot+leads may be violated in the process of cotunneling unlike the
rotation spin symmetry [see the terms $\sim J_{3,5}$ in the
Hamiltonian (\ref{SWE})].

Some of these constants are positive $(J_{1,4,6})$ like in
conventional SW transformation, and some are negative
$(J_{2,3,5})$. Thus the Kondo effect arises as a result of
interplay of antiferromagnetic and ferromagnetic exchange
involving both spin and pseudospin variables. This interplay is
described by scaling equations obtained in a framework of
Anderson's "poor man scaling" procedure \cite{Andrg}. The system
of scaling equations describes the flow of exchange vertices, with
reducing the energy scale $\bar D$ from the boundary value $\bar
D_{SW}$ to the Kondo temperature $T_K$. It has the following form:
\begin{eqnarray}
 \frac{dj_1}{dt} &=&
 -\left[j_1^2+\frac{j_4^2}{2}+j_4j_6+\frac{j_5^2}{2}\right],
\nonumber \\
 \frac{dj_2}{dt} &=&
 -\left[j_2^2+\frac{j_4^2}{2}-j_4j_6+\frac{j_5^2}{2}\right],
 \nonumber \\
 \frac{dj_3}{dt} &=&
 -\left[j_3^2+j_5^2\right],
 \label{scal} \\
 \frac{dj_4}{dt} &=&
 -\left[j_4(j_1+j_2+j_6)+j_6(j_1-j_2)\right],
 \nonumber\\
 \frac{dj_5}{dt} &=&
 -\frac{j_5}{2}\left[j_1+j_2+j_3-j_6\right],
 \nonumber \\
 \frac{dj_6}{dt} &=&
 -j_6^2. \nonumber
 \end{eqnarray}
Here $j_i=\rho_0 J_i$, and the scaling variable is $t=\ln \bar D$.

Analysis of solutions of the scaling equations (\ref{scal}) with
initial values of coupling parameters listed in Eq. (\ref{J-E}),
shows that the symmetry-breaking vertices $j_{3,5}$ are
irrelevant, and the vertex $j_2$, which is negative at the
boundary $\bar D =\bar D_{SW}$ evolves into positive domain and
eventually enters the Kondo temperature. The latter is given by
the following equation
\begin{equation}\label{kondoe}
T_k={\bar
D}\exp\left\{{-\frac{2}{j_1+j_2+\sqrt{2}j_4+2j_6}}\right\}.
\end{equation}
We see from this equation that both spin and pseudospin coupling
contribute to the Kondo cotunneling on equal footing, so that
$T_K$ is enhanced due to involvement of additional degrees of
freedom connected with electron permutations in a triangle.

The tunnel contact reduces the symmetry of TTQD when the
conditions (\ref{oneax}) are satisfied. In this case the double
orbital degeneracy of the level $E_{DE}$ is removed by tunnel
coupling. Then the generic symmetry of TTQD is usual $SU(2)$
symmetry, but an accidental degeneracy between the levels
$E_{DE(+)}$ and $E_{DA_1}$ arises under certain condition, and in
this case the $SU(4)$ symmetry is restored. This situation as well
as the properties of TTQD with ${\cal N}=2$ will be considered in
more detailed forthcoming publication.

\section{Conclusions}

Dynamical symmetry is a universal tool, which allows to derive the
effective spin Hamiltonians describing low-energy cotunneling with
spin reversal through composite quantum dots. These dots possess
both the continuous rotation symmetry in spin space and discrete
rotation symmetry in real space. We considered in this paper the
relatively simple example of symmetric triangular quantum dot in
contact with metallic leads in a specific geometry where the dot
imposes its discrete rotation symmetry on the electrons in the
leads. The tunnel problem is mapped on the Coqblin-Schrieffer (CS)
model of magnetic impurity with "orbital" degrees of freedom. Like
in the latter case, the symmetry of the Kondo center is $SU(2n)$.
The parameter $n$ characterizes additional orbital degeneracy in
conventional CS model. In case of artificial triangular molecule
it describes the pseudospin operator of finite
clockwise/anti-clockwise rotations of TTQD in the process of
electron cotunneling.

\end{document}